# Remote Stimulated Triggering of Quantum Entangled Nuclear Metastable States of $^{115m}$In

D. L. Van Gent, Nuclear Science Center, Louisiana State University, Baton Rouge, USA

Abstract

We report experiments in which two indium foils were quantum entangled via photoexcitation of stable $^{115}$In to radioactive $^{115m}$In by utilizing Bremsstrahlung gamma photons produced by a Varian Compact Linear Accelerator (CLINAC). After photo-excitation, remote triggering of the "master" foil with low energy gamma photons, yielded stimulated emissions of 336 keV gamma photons from quantum entangled $^{115m}$In in the "slave" foil located up to 1600 m away from the "master" foil. These experiments strongly demonstrate that useful quantum information can be transferred through quantum channels via modulation of quantum noise (accelerated radioactive decay of $^{115m}$In metastable nuclei). Thus, this modality of QE transmission is fundamentally different from optical QE information transfer via quantum entangled space "q-bits" as developed by information theorists for quantum channel information transfer. Additionally, there is no obvious potential for signal degradation with increasing distance nor the problems associated with misalignment of optical information transfer systems

1. **Introduction**

The possibility of instantaneous transfer of quantum information over macroscopic distances was first alluded to by Einstein, Podolsky, and Rosen [1]. They wrote with strong conviction that General Relativity and QED are fundamentally at odds with each other in this respect, since QED seems to indicate the possibility of "instantaneous" transfer of quantum information over long distances. According to QED theory, it should be possible to send quantum-encoded (polarized) photons through optical transfer media, allowing instantaneous transfer of quantum information in direct contradiction to General Relativity (GR). Currently, Information Theory experts generally agree that it is doubtful that useful information can be transmitted faster than light via QE photons [2], but it is widely acknowledged that, in theory, quantum noise can be transferred instantaneously to any point in the universe via QE systems.

Several experiments carried out in the last decade strongly demonstrate the validity of quantum entanglement of photons over macroscopic distances, most recently at 100 km. In 2003, Andrew Shields [3] and his colleagues at Toshiba Research Europe Ltd. (Cambridge, UK) carried out quantum cryptography experiments by encoding information in the polarization of individual photons sent over 100 km of optical fiber, breaking an earlier record by about 40 km. However, for a variety of reasons, photons are less and less likely to be detectable the farther they travel.

Difficulties that must be overcome if such optically-based information transfer technologies are to become practical and commercially viable are common to all optical communication modalities, such as requisite precise optical system alignment, accurate timing necessary for encoded photon packet reception, photon signal degradation, and environmental zero point vacuum flux induced de-coherence of "unprotected" quantum entangled systems over distance and time (O'Connel, 2002)[4]. Optical technologies developed for this purpose thus far allow for only very limited applications of quantum channel transmissions.

We have previously reported [5] strong evidence that high-energy gamma and Bremsstrahlung quantum entangled photons can be transferred for extended periods of time into nuclear radioactive metastable nuclear states of certain photo-excited metals. Paired QE nucleonic metastable states must conform to quantum spin and angular momentum conservation laws even when separated by macroscopic distances similar to QE paired photons.

The relatively new field of study pertaining to nuclear photon pumping into metastable nucleii and subsequent direct "triggering" for release of gamma photon energy of isomers has been coined, "Nucleonics" [6] being essentially the nuclear analog to the field of "electronics." DOE funded work is currently ongoing with the desired end result being the storage and release of Giga joules/gm of the most promising isomer $^{178m2}$Hf with a half-life of gamma decay of 31 years [7].

**2. Methodology**

It is well known that low energy photon pairs from atomic radiative cascade are entangled [8]. In the experiments reported here, entangled gamma photons were produced from both radio-isotopic $^{60}$Co nuclear decay and CLINAC Compact Linear Accelerator indium metal foil irradiation. The quantum entanglement of high energy gamma and Bremsstrahlung photons can be transferred via nucleonic photon pumping of metastable nuclei.

In this experiment, two identical 5x5 cm 0.25 mm thick 99.999% pure natural indium were photo-excited for various lengths of time together aligned in the same plane either in the HICS irradiation chamber for 30 hours or with the CLINAC accelerator Bremsstrahlung beam for 20 minutes. Figure 1 depicts a conceptual rendering of the experimental design for CLINAC irradiation of the indium foils.

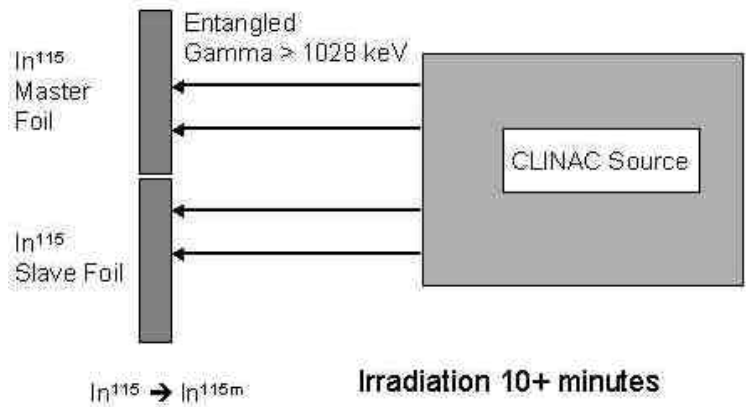

Figure 1.  CLINAC photo-excitation of two indium foils

During gamma counting of the resulting photo-excited indium foils with a Canberra high purity intrinsic germanium gamma spectrometer interfaced with a multichannel analyzer, we observed a direct correlation between two entangled foils during spontaneous decay.  Further investigation revealed that direct triggering of one of the QE paired foils with low-energy gamma photons resulted in an indirect correlated emission from the second foil located at a distance of at least 12 meters and separated by 15 cm of lead. The experimental design is depicted in Figure 2.

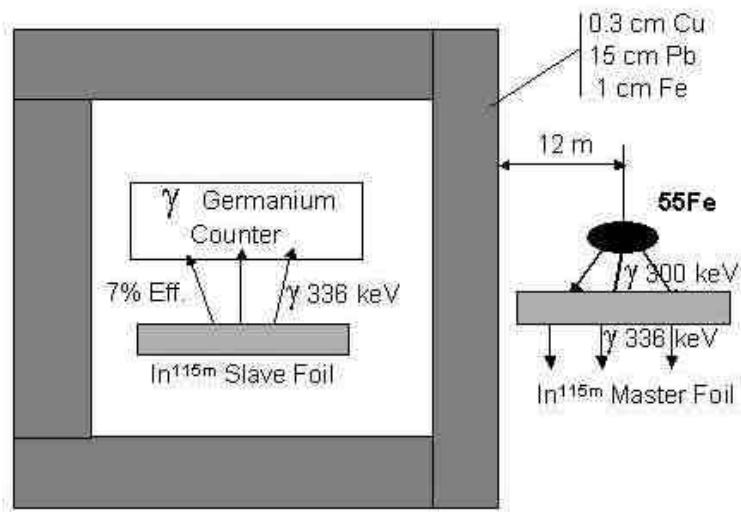

Figure 2.  Experimental design of remote gamma triggering with two indium foils

## 3.  Results

During the remote triggering experiments, HICS irradiated indium foils yielded statistically insignificant remote gamma triggered photons because HICS irradiated indium foils evidence QE $^{115m}$In states of only 3.5% QE doublets. Therefore, we did not attempt remote triggering experiments with HICS irradiated indium foils.

CLINAC irradiated indium foils yielded statistically significant remote triggered gamma photons based on the fact that 9% of metastable $^{115m}$In are QE doublets and 9% are QE triplets.

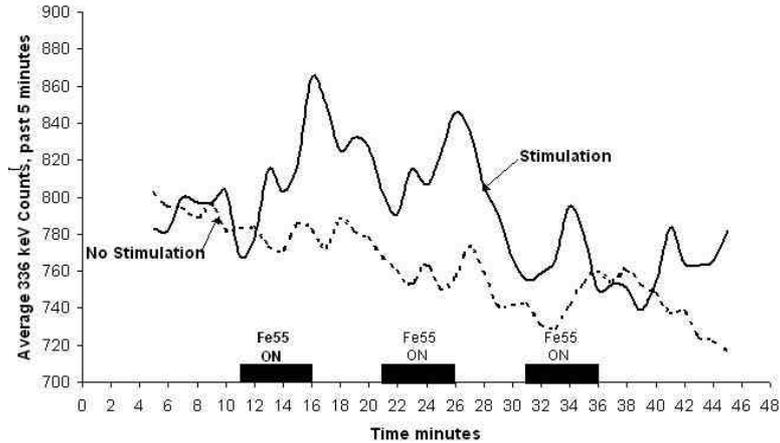

Figure 3. Graph of remote triggering of CLINAC photo-excited indium foils at 12 meters as compared to CLINAC photo-excited indium foils with no stimulation. Germanium counter.

A typical last five minute running average of one minute gamma counts of the "slave" foil is depicted in Figure 3. The same data has been used in Figure 4 to outline the results. The calculated average for each interval is shown for the entire interval.

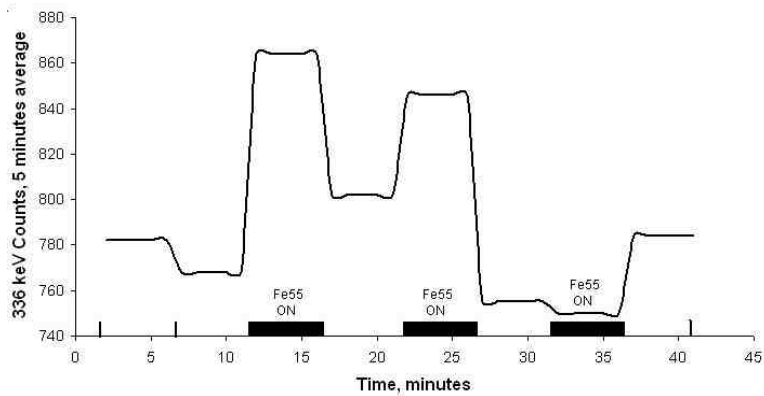

Figure 4. Graph of remote triggering of CLINAC photo-excited indium foils at 12 meters. Averages shown during the various time intervals. Germanium counter.

The experiment was repeated again at 12 meters between the master foil and the slave foil using a NaI counter. The results are depicted in Figure 5 using the average recorded in each interval and showing for the whole interval.

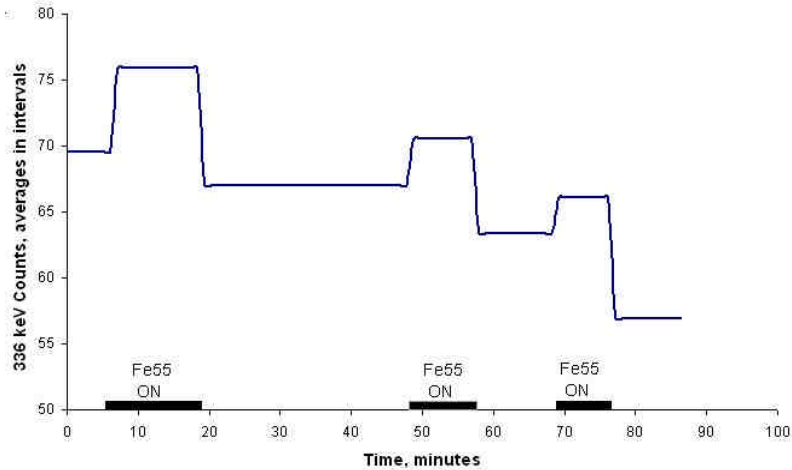

Figure 5. Graph of remote triggering of CLINAC photo-excited indium foils at 12 meters. Averages shown during the various time intervals. NaI counter.

Another experiment was conducted with a distance of 1600 meters separating the master foil and the slave foil with the same results.

It clearly demonstrates that remote triggering of the "master" foil resulted in a 4-sigma above the $^{115m}$In spontaneous decay baseline of 336 keV characteristic gamma photon emission from the "slave" foil as measured by the gamma counting system. The two foils were separated by at least 12 meters, then 1600 meters, and 15 cm of lead. It is apparent that it is possible to stimulate "master" foil multiple times, however, it appears that only a limited number of QE states are available for remote triggering.

**4. Conclusion**

This experiment strongly demonstrates that useful quantum information can be transferred through quantum channels via modulation of quantum noise (accelerated radioactive decay of $^{115m}$In). Thus, this modality of QE transmission is fundamentally different from optical QE information transfer via quantum entangled space "q-bits" as developed by information theorists for quantum channel information transfer. Additionally, there is no obvious potential for signal degradation with increasing distance nor the problem of misalignment of optical information transfer systems.

Although $^{115m}$In metastable states have a spontaneous decay half-life of 4.68 hours, other much longer-lived metastable states such as $^{178m2}$Hf with a half-life of 31 years could potentially be utilized for viable global communications.

Even though only two foils were quantum entangled per irradiation during this experiment, there is no foreseeable reason why multiple numbers of foils could not be utilized as well. If this is possible, one "master" foil could be utilized to remotely trigger multiple QE "slave" foils.

**Acknowledgement**

I thank Professor Robert Desbrandes (LSU emeritus from Petroleum Engineering) for his help in the experiments and their interpretation. I also thank the Veterinary School of LSU for the use of its CLINAC accelerator and the Nuclear Science Center of LSU for using its HICS Cobalt 60 source and Germanium gamma counter